\documentclass{raa}            % referee version: for submission

\usepackage{graphicx,times}             %for PS/EPS graphics inclusion, new
\usepackage{natbib}
\usepackage{amssymb,amsmath}
\bibpunct{(}{)}{;}{a}{}{,}

\usepackage[a4paper=true,dvipdfm=true,pagebackref=true]{hyperref}
\hypersetup{colorlinks = true, linkcolor = green, anchorcolor = red, citecolor = blue, filecolor = red, pagecolor = red, urlcolor = red}

\begin{document}

	\title{Investigation of the parameters of spiral pattern in galaxies: the arm width}
   %\title{Investigation of the parameters of spiral pattern in galaxies based on multiband photometry: arm width}
   %\title{Investigation of the arm width in spiral galaxies}
%   \subtitle{I. Place Your Subtitle Here}

   \volnopage{Vol.0 (20xx) No.0, 000--000}      %%preserved for Editor. DOn't remove!
   \setcounter{page}{1}          %%starting page, preserved for Editor. DOn't remove!

   \author{Aleksandr Mosenkov
      \inst{1, 2}
   \and Sergey Savchenko
      \inst{3, 4}
   \and Alexander Marchuk
      \inst{3}
   }

   \institute{Department for Management of Science and Technology Development, Ton Duc Thang University, Ho Chi Minh City, Vietnam; {\it aleksandr.mosenkov@tdtu.edu.vn}\\
     \and
     Faculty of Applied Sciences, Ton Duc Thang University, Ho Chi Minh City, Vietnam\\
     \and
     Saint Petersburg State University, Department of Astrophysics, St. Petersburg, 198504 Russia\\
     \and
     Special Astrophysical Observatory, Russian Academy of Sciences,
     Nizhnii Arkhyz, 369167 Russia\\
\vs\no
   {\small Received~~20xx month day; accepted~~20xx~~month day}}

\abstract{In this work we determine the parameters of spiral structure for a sample of face-on spiral galaxies. In practice, the solution of this problem is a hard task because of the diversity of the observed characteristics of spiral structure, such as the arm number, their shape, arm contrast etc. In this work we study spiral structure in galaxies based on an analysis of photometric cuts perpendicular to the arm direction. The method is based on an approximation of these slices with an analytical function and derivation of the parameters of spiral structure (arm width, asymmetry, pitch angle) using the fitted parameters of this approximation. The algorithm has been applied to a sample of 155 galaxies selected from the Sloan Digital Sky Survey in different passbands. In this paper we only consider the results on the arm width: most spirals show an increase of their width with galactocentric distance. Only 14 per cent of galaxies in our sample show an opposite trend or have an almost constant arm width at all radii.
}

   \authorrunning{A. Mosenkov, S. Savchenko \& A. Marchuk }            %author_head in even pages
   \titlerunning{Parameters of spiral pattern in galaxies }  % title_head in odd pages

   \maketitle
%% The author head (on even pages) and the title head (on odd pages) will be
%% automatically extracted from \author{} and \title{}. Whenever the title is too long,
%% you will be asked to supply a shorter one by inserting either \authorrunning{} or
%% \titlerunning{} before \maketitle. Anyway, you can specify your own heads.
%%
%%
%% Note: In the following text body of your manuscript, please note several differences from
%%       other major journals:
%% (1) \subsection{Please Capitalize the First Letter of Each Notional Word in Subsection Title}
%% (2) Please Capitalize the First Letter of Each Notional Word in all tables' captions

%
%________________________________________________ sections below
%
\section{Introduction}           %% first-level sections will be auto-capitalized
\label{sect:intro}
Spiral galaxies are ubiquitous in the Local Universe \citep{2006MNRAS.373.1389C}. The fact that we observe a prominent spiral structure in so many disc galaxies should be explained. If all the spirals
were material, they would be tightly wound up by the differential rotation of the disc and decay after several galactic years which contradicts our observations (the so-called winding problem). Other proposed mechanisms to explain this phenomenon, however, are still under debate \citep[see the review by][]{2014PASA...31...35D}. Numerical simulations support the dynamical spirals which are swing amplified by the differential rotation of the disc \citep[see e.g.][]{1966ApJ...146..810J,1978ApJ...223..129G}. An alternative approach is the quasi-stationary wave theory which explains the existence of grand-design galaxies with two distinct spiral arms very well \citep{1964ApJ...140..646L,1989ApJ...338...78B}. In addition to that, spirals can be induced by tidal interactions \citep{1969ApJ...158..899T,1972ApJ...178..623T} or be bar-driven \citep{1996A&A...309..381K,2009MNRAS.394.1605H}. It is also believed, that different mechanisms can be dominant under different circumstances. Despite the long history of the research of spiral structure in galaxies, the problem of its formation seems far from being solved as many collected observational facts cannot give us a certain answer on the question of what is the dominant mechanism among all these alternatives to form spiral arms in galaxies.

In this brief paper, we investigate the width of spiral structure in the optical, in the framework of our project aimed at studying spiral structure in galaxies in a wide range of wavelengths \citep{Savchenko}. To our surprise, this kind of study was done only for several galaxies \citep{2015ApJ...800...53H} and our own Milky Way \citep{2014ApJ...783..130R}, but based on the positions of H{\sc ii} regions
and  masers. Except for these two studies, we found no other papers where the width of spiral arms would be considered as a parameter which characterises spiral structure in galaxies. We decided to fill in this gap and initiated a detailed study of spiral structure in galaxies. In this paper we will only consider results on the arm width.

Our paper is organised as follows. In Sect.~\ref{sect:data} we briefly describe our sample and data preparation.  In Sect.~\ref{sect:analysis} we present our method for fitting spiral structure in galaxies. In Sect.~\ref{sect:results} we present main results of our study regarding the width of spiral arms and summarize main conclusions in Sect.~\ref{sect:conclusion}.

\section{Sample and Data}
\label{sect:data}
Our sample is described in detail in \citet{Savchenko}. To create a sample of spiral galaxies, we used the EFIGI \citep{2011A&A...532A..74B} and GalaxyZoo \citep{2008MNRAS.389.1179L} samples. From these samples we initially selected rather large spiral galaxies with the optical diameter $d25>50$'' (at the isophote 25\,mag\,arcsec$^{-2}$ in the $B$ band calculated from the parameter $logd25$ from HyperLeda, \citealt{2014A&A...570A..13M}). These galaxies were inspected by eye several times to exclude interacting and highly-inclined galaxies. Also, we rejected galaxies with a quite unclear, too flocculent spiral structure where no individual spirals arm can be traced well. Our final sample consists of 155 galaxies of Sa--Sd types. The average distance to the sample galaxies is $90.0\pm57.7$~Mpc.

In our study we use the three $gri$ bands from the the Sloan Digital Sky Survey; the images were taken from the DR13 \citep{2017ApJS..233...25A}. All the images were reduced for the purposes of this study: sky-subtracted, cropped, and de-projected. The de-projection was done using the measured position angles and flattenings of the outermost galaxy isophotes which were estimated from isophote fitting. We also subtracted the azimuthally averaged profiles from the real images to suppress the axisymmetric components of the galaxy and to enhance the contrast of the spiral structure.
Finally, we extracted a point spread function (PSF) image for each galaxy frame.

\section{Data Analysis}
\label{sect:analysis}
\begin{figure}[h]
  \begin{minipage}[t]{0.495\textwidth}
  \centering
   \includegraphics[width=65mm, clip=True, trim=0cm 1cm 0 1cm]{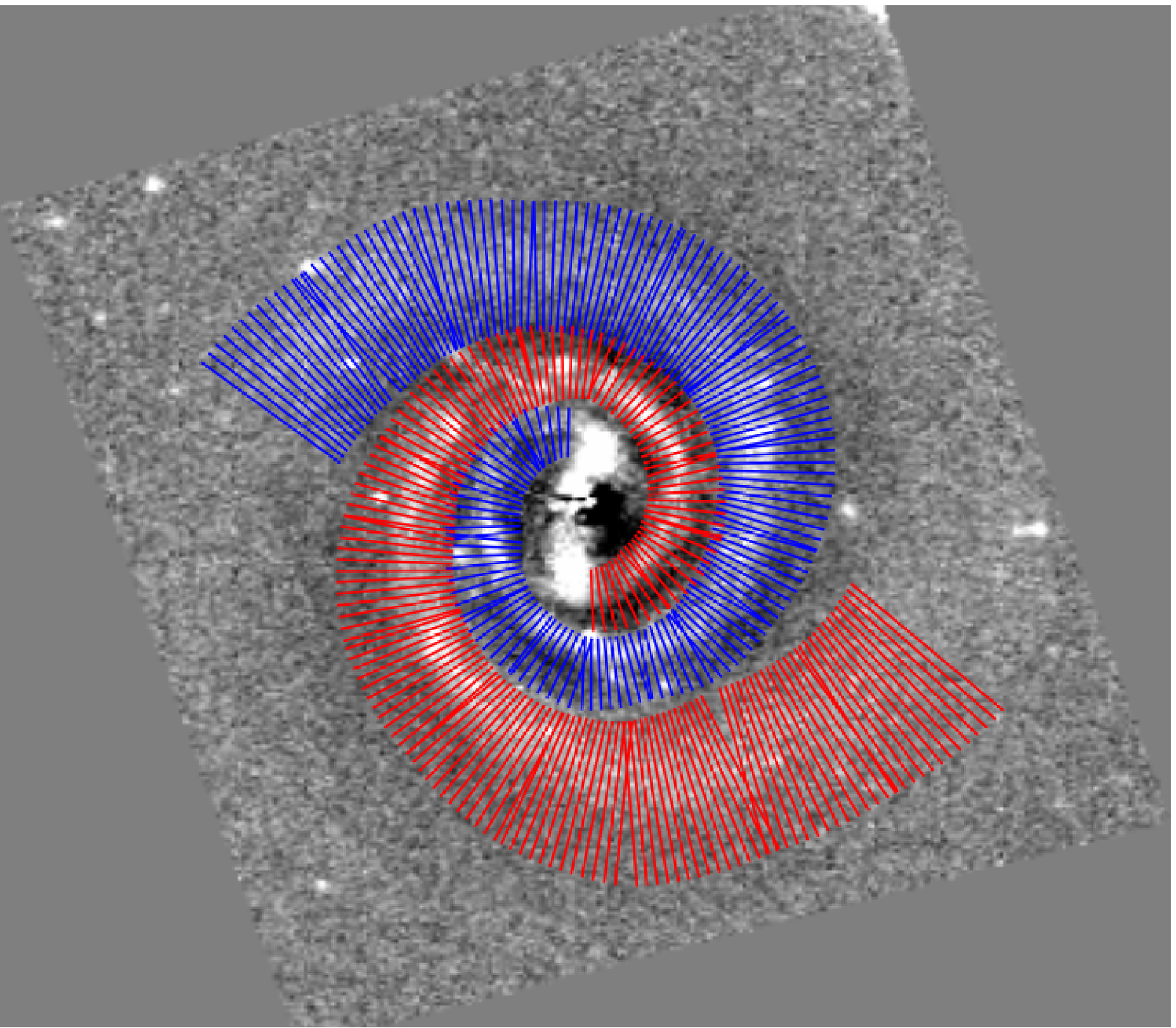}
  \caption{{\small A set of slices made perpendicular to the arm\\ structure of PGC~2182. Different colours show slices\\ made for
    different arms.}}
  \label{Fig:Fig1}
  \end{minipage}%
 \begin{minipage}[t]{0.495\linewidth}
  \centering
   \includegraphics[width=70mm, clip=True, trim=0cm 0.25cm 0 0.25cm]{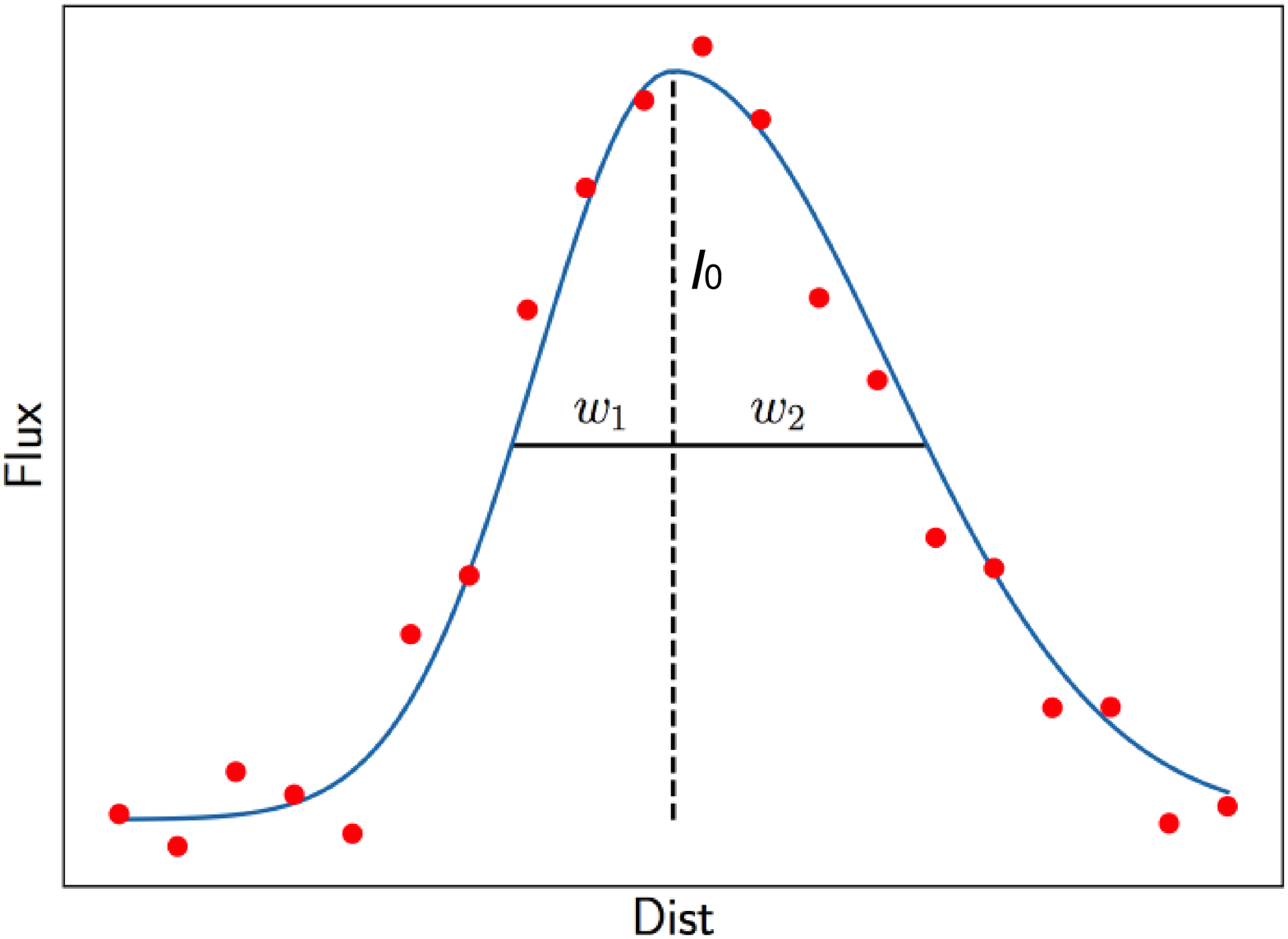}
   \caption{{\small A schematic view for a single photometric cut \\
       and its fit by an asymmetric gaussian function. The red\\ dots are data points, the solid line is a best fit.} }
  \label{Fig:Fig2}
  \end{minipage}%
\end{figure}

We give details of our data analysis in \citet{Savchenko}. For each galaxy we marked at least one spiral arm, which can be traced in the de-projected residual image we created in Sect.~\ref{sect:data}. We trace the arm from the ends of a bar or beyond a bulge, if these components are present in the galaxy, to the galaxy outskirts at outermost galaxy isophotes. Each spiral arm was traced using `circle' regions from the ds9 program\footnote{\url{http://ds9.si.edu/site/Home.html}}: these regions were placed manually along the central line of the arm, estimated approximately by eye, with a circle radius extending till the middle of the inter-arm space on either side of the arm inwards and outwards.
Then a special algorithm was used to make a set of slices perpendicular to the arm in a number of points with a length of each slice to be equal to the diameter of the circle
regions (see Fig.~\ref{Fig:Fig1}).
 Then we fit each slice with an asymmetric Gaussian function of distance from the galaxy center $r$ with four free parameters, the central flux $I_0$, the peak location $r_{\mathrm{peak}}$, and the two half-widths $w_1$ and $w_2$ (see Fig.~\ref{Fig:Fig2}): 

\begin{equation}
  \label{Eq:Eq1}
I_{\mathrm{model}}(r) = I_0 \times
  \begin{cases}
    \exp \left( - \frac{ \left[ r - r_{\mathrm{peak}} \right]^2 }{w_1^2} \right), & r < r_{\mathrm{peak}} \\
    \exp \left( - \frac{ \left[ r - r_{\mathrm{peak}} \right]^2 }{w_2^2} \right), & r > r_{\mathrm{peak}}\,.
  \end{cases}
\end{equation}

The full width of the arm at a certain point can be calculated as $w=w_1+w_2$. During the fitting, we convolve the model of the asymmetric gaussian with the created PSF to take into account the image smearing by the atmosphere and optics. When the fitting is done
for each slice, we have a measurement of the arm width at a number of points along the arm.

\section{Results}
\label{sect:results}

Obviously, from this fitting we can estimate many other parameters of the spiral structure (and in the other passbands), as, for example, the pitch angle and its variation along the radius, the fraction of the spiral arm to the total galaxy luminosity, and arm colours. Studying these important parameters of the spiral structure is beyond the scope of this paper. We present all the results in \citet{Savchenko}.

\begin{figure}[h]
  \begin{minipage}[t]{0.495\linewidth}
  \centering
   \includegraphics[width=75mm]{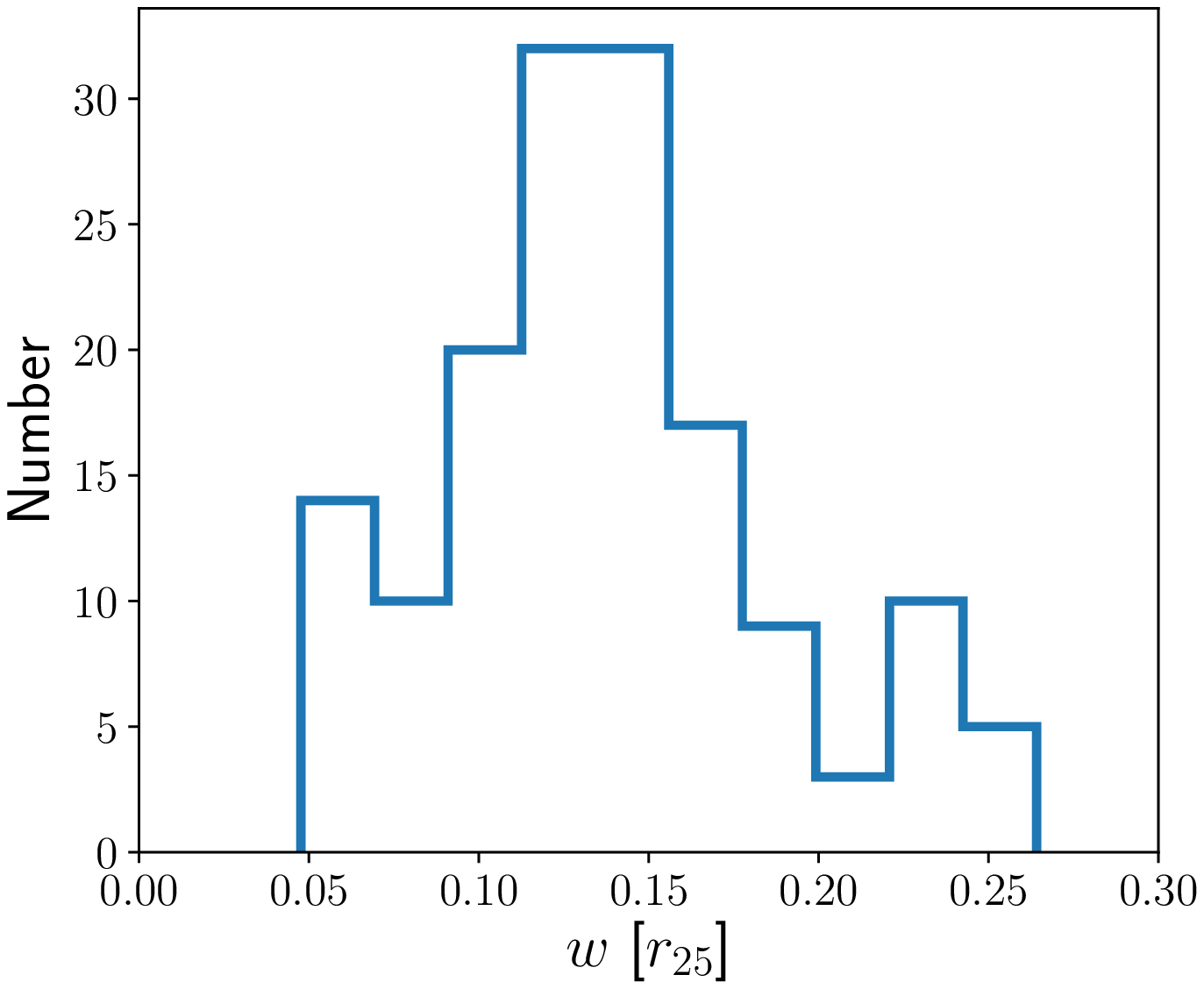}
   \caption{{\small Distribution by the average arm width for the \\ sample galaxies.} }
  \label{Fig:Fig3}
  \end{minipage}%
  \begin{minipage}[t]{0.495\textwidth}
  \centering
   \includegraphics[width=75mm]{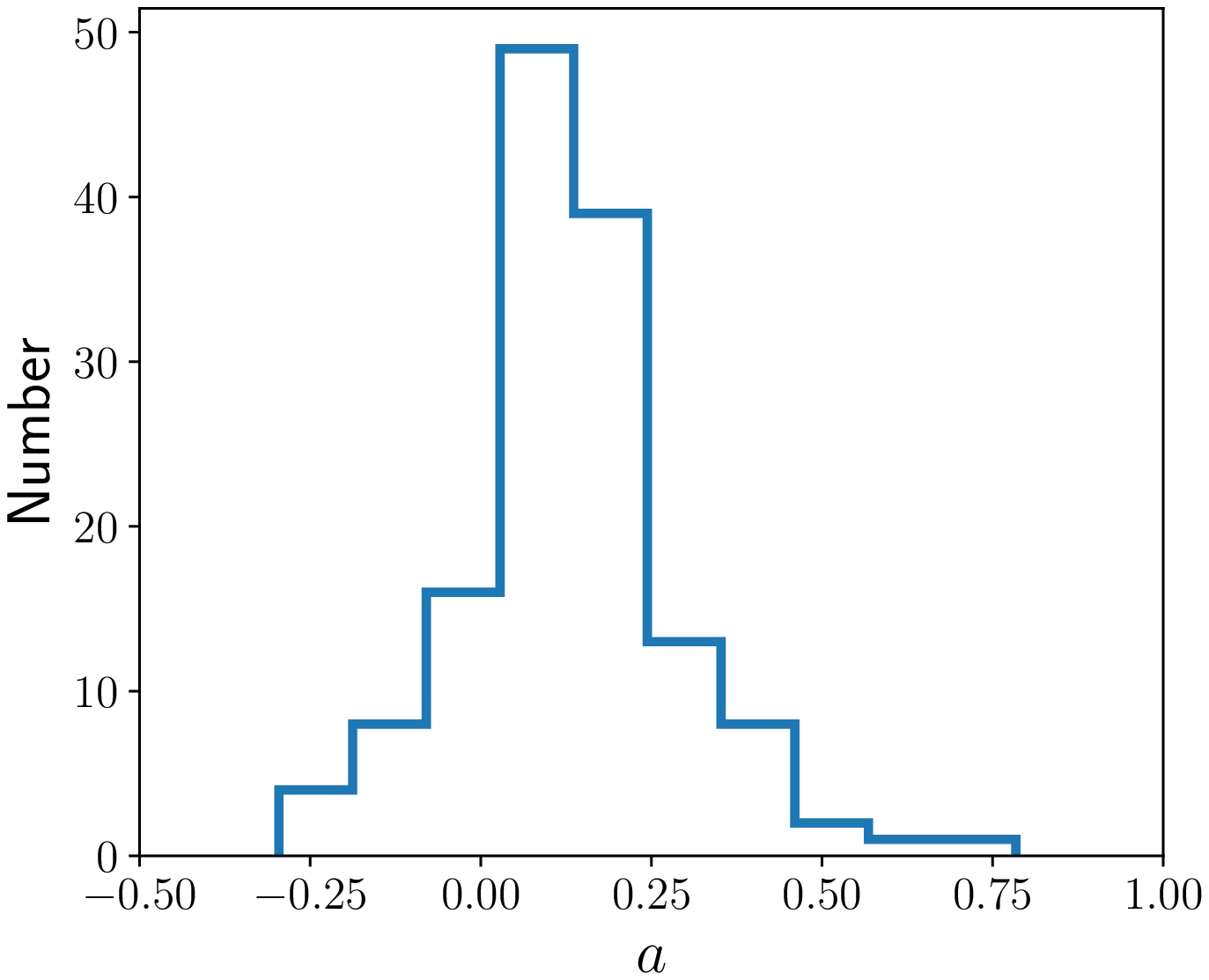}
  \caption{{\small Distribution by the slope $a$ for the dependence \\ between the local arm width and galactocentric distance.}}
  \label{Fig:Fig4}
  \end{minipage}%
\end{figure}

\begin{figure}
   \centering
   \includegraphics[width=10cm]{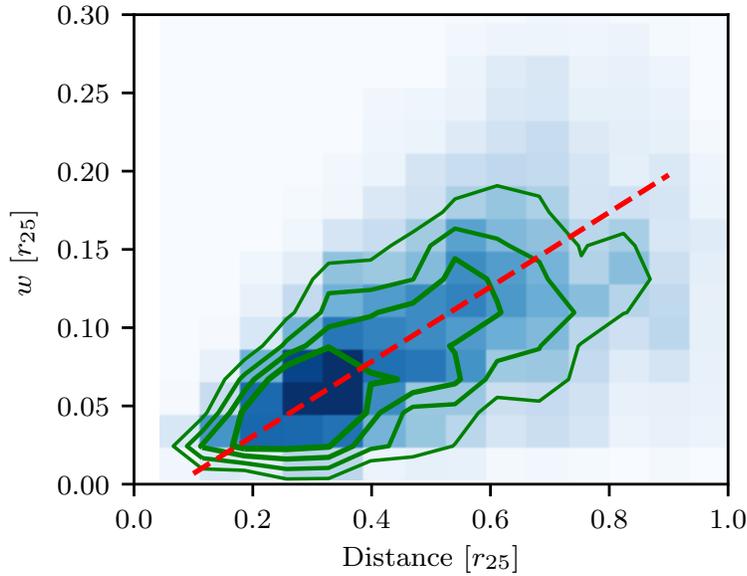}
   \caption{Dependence of the arm width with radius (normalized by the optical radius) created for the whole sample. The darker colour corresponds to a higher concentration of the points, with the green overlayed isolevels. The red dashed line shows the linear regression with the slope $0.239$.}
   \label{Fig:Fig5}
   \end{figure}

\begin{figure}
   \centering
   \includegraphics[width=15cm]{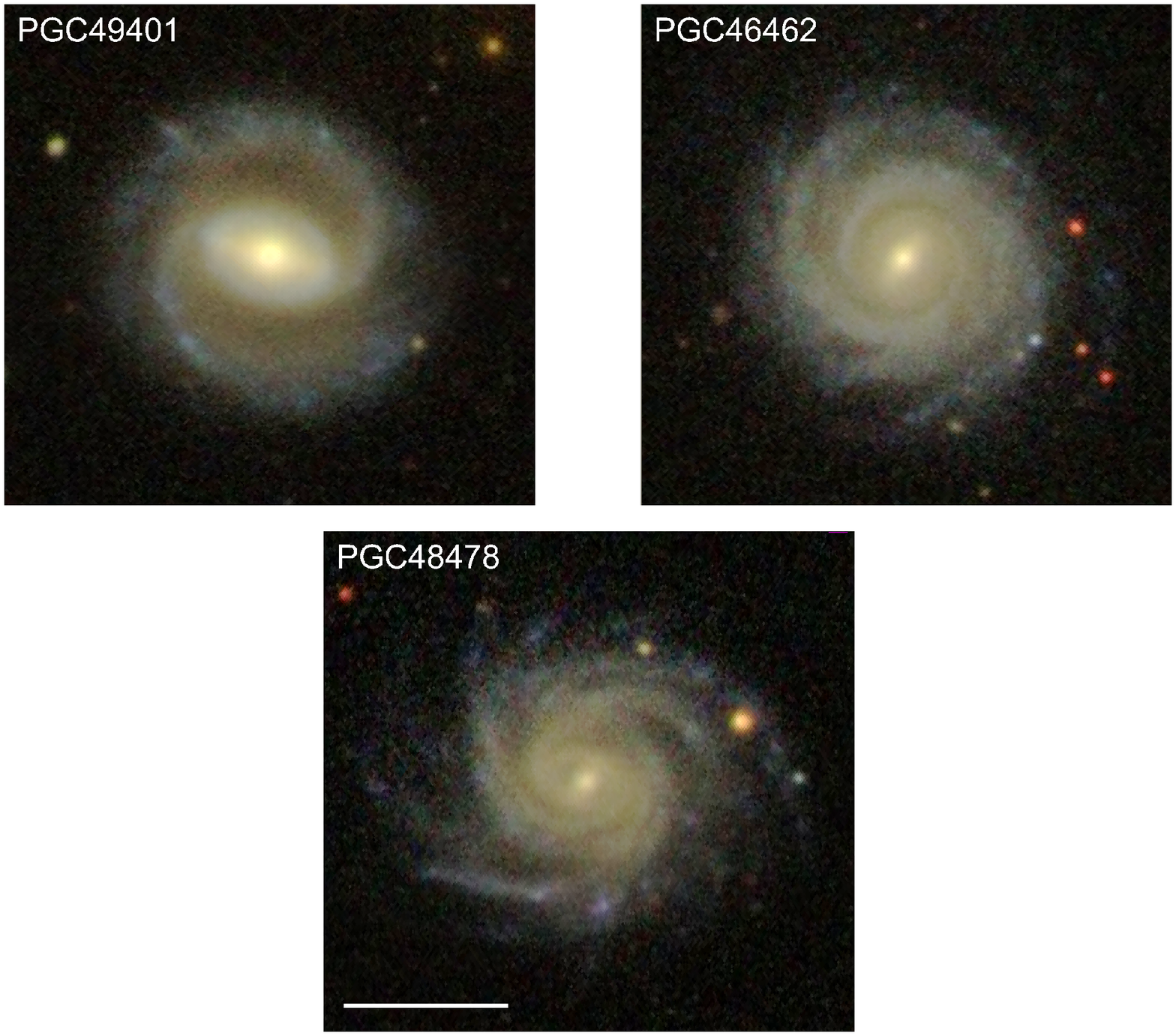}
   \caption{Examples of galaxies with different behaviour of the arm width with radius. The galaxy PGC\,49401 has the spiral arms with the largest slope $a=0.78$, the galaxy PGC\,46462 has $a=0.0$, and the arms in PGC\,48478 show the apparent decrease of the arm width with radius ($a=-0.30$). The white bar depicts 30~arcsec.}
   \label{Fig:Fig6}
   \end{figure}

In Fig.~\ref{Fig:Fig3} we show the distribution of the mean arm width for the sample galaxies. The averaging was done 1) along the radius for each arm, and 2) for all arms in a galaxy. The mean arm width for the whole sample is $0.14 \pm 0.05$ in units of 25-th isophote radius $r_{25}$ in the $r$ band.
Most importantly, we can study how the arm width changes with radius in each galaxy. To quantify the radial variation of the spirals, we fit a linear
regression into the radius--width relation, and see the slope $a$ of the linear fit. The distribution of our sample galaxies by the slope value is shown
in Fig.~\ref{Fig:Fig4}. From this figure we can clearly see that for most galaxies (86 per cent) in our sample the arm width increases with galactocentric distance
(they have positive values of $a$). Galaxies from \citet{2015ApJ...800...53H} demonstrate the width slope values ranging from $a=0.05$ to $a=0.12$, which all lie within the central histogram peak in Fig.~\ref{Fig:Fig4}.
 Fig.~\ref{Fig:Fig5} shows the radius--width relation for all arms of all galaxies of our sample as a density map with a fit by a linear function $w = a \cdot r_{25} + b$ for the whole sample, where the slope
 $a$ was found to be equal to 0.24, and the intercept $b=-0.02$.
 However, there is some number of galaxies in our sample, which show almost constant or even narrower arm width at larger radii.
 In Fig.~\ref{Fig:Fig6}, we show snapshots for three example galaxies in our sample, which exhibit different behaviour of the arm width with radius.

Our results agree well with the results from \citep{2014ApJ...783..130R,2015ApJ...800...53H}: in most sample galaxies we observe that the arm width grows with radius. The swing amplification theory can explain this behaviour \citep{1981seng.proc..111T,2015ApJ...800...53H}, whereas for the other theories (see Sect.~\ref{sect:intro}) this should be studied. The question on why some galaxies do not show an increase of the arm width should be investigated thoroughly in a subsequent study. Here we can only speculate that this can be related to a reduced star formation (truncation of the H{\sc i} profile) at the periphery of these galaxies.

\section{Conclusions}
\label{sect:conclusion}
We have studied a sample of face-on non-interacting spiral galaxies in order to retrieve the parameters of their spiral structure. One of the advantages of this work is that we studied the arm width as one of the important parameters of spiral structure. Using a specially developed method for fitting the spiral arms in galaxies, we were able to measure the arm width and its change with radius. We found that most spirals in our sample (86 per cent) show a growing arm width with radius. This is in line with the physical essence of the swing amplification mechanism for explaining the phenomenon of the spiral structure in galaxies.
In our next studies we are about to study the other retrieved parameters of the spiral structure in great detail and link them to different proposed models for generating spiral structure in galaxies.

\begin{acknowledgements}
The work was supported by the RFBR grant 18-32-00194.
Funding for the SDSS has been provided by the Alfred P. Sloan
Foundation, the Participating Institutions, the National Science
Foundation, the US Department of Energy, the National Aeronautics
and Space Administration, the Japanese Monbukagakusho, the
Max Planck Society, and the Higher Education Funding Council for
England. The SDSS Web Site is \url{http://www.sdss.org/}.
This research makes use of the NASA/IPAC Extragalactic Database (NED) which is operated by the Jet Propulsion Laboratory, California Institute of Technology, under contract with the National Aeronautics and Space Administration, and the LEDA database (\url{http://leda.univ-lyon1.fr}).

\end{acknowledgements}

\bibliographystyle{raa}
\bibliography{article}

\begin{thebibliography}{18}
\providecommand\natexlab[1]{#1}
\providecommand\JournalTitle[1]{#1}

\bibitem[{Albareti} {et~al.}(2017)]{2017ApJS..233...25A}
{Albareti}, F.~D., {Allende Prieto}, C., {Almeida}, A., {et~al.} 2017, \apjs,
  233, 25

\bibitem[{Baillard} {et~al.}(2011)]{2011A&A...532A..74B}
{Baillard}, A., {Bertin}, E., {de Lapparent}, V., {et~al.} 2011, \aap, 532, A74

\bibitem[{Bertin} {et~al.}(1989)]{1989ApJ...338...78B}
{Bertin}, G., {Lin}, C.~C., {Lowe}, S.~A., \& {Thurstans}, R.~P. 1989, \apj,
  338, 78

\bibitem[{Conselice}(2006)]{2006MNRAS.373.1389C}
{Conselice}, C.~J. 2006, \mnras, 373, 1389

\bibitem[{Dobbs} \& {Baba}(2014)]{2014PASA...31...35D}
{Dobbs}, C., \& {Baba}, J. 2014, \pasa, 31, e035

\bibitem[{Gerola} \& {Seiden}(1978)]{1978ApJ...223..129G}
{Gerola}, H., \& {Seiden}, P.~E. 1978, \apj, 223, 129

\bibitem[{Harsoula} \& {Kalapotharakos}(2009)]{2009MNRAS.394.1605H}
{Harsoula}, M., \& {Kalapotharakos}, C. 2009, \mnras, 394, 1605

\bibitem[{Honig} \& {Reid}(2015)]{2015ApJ...800...53H}
{Honig}, Z.~N., \& {Reid}, M.~J. 2015, \apj, 800, 53

\bibitem[{Julian} \& {Toomre}(1966)]{1966ApJ...146..810J}
{Julian}, W.~H., \& {Toomre}, A. 1966, \apj, 146, 810

\bibitem[{Kaufmann} \& {Contopoulos}(1996)]{1996A&A...309..381K}
{Kaufmann}, D.~E., \& {Contopoulos}, G. 1996, \aap, 309, 381

\bibitem[{Lin} \& {Shu}(1964)]{1964ApJ...140..646L}
{Lin}, C.~C., \& {Shu}, F.~H. 1964, \apj, 140, 646

\bibitem[{Lintott} {et~al.}(2008)]{2008MNRAS.389.1179L}
{Lintott}, C.~J., {Schawinski}, K., {Slosar}, A., {et~al.} 2008, \mnras, 389,
  1179

\bibitem[{Makarov} {et~al.}(2014)]{2014A&A...570A..13M}
{Makarov}, D., {Prugniel}, P., {Terekhova}, N., {Courtois}, H., \& {Vauglin},
  I. 2014, \aap, 570, A13

\bibitem[{Reid} {et~al.}(2014)]{2014ApJ...783..130R}
{Reid}, M.~J., {Menten}, K.~M., {Brunthaler}, A., {et~al.} 2014, \apj, 783, 130

\bibitem[{Savchenko} {et~al.}(2020)]{Savchenko}
{Savchenko}, S., {Marchuk}, A., {Mosenkov}, A., \& {Grishunin}, K. 2020, arXiv
  e-prints, arXiv:2001.09110

\bibitem[{Toomre}(1969)]{1969ApJ...158..899T}
{Toomre}, A. 1969, \apj, 158, 899

\bibitem[{Toomre}(1981)]{1981seng.proc..111T}
{Toomre}, A. 1981, in Structure and Evolution of Normal Galaxies, ed. S.~M.
  {Fall} \& D.~{Lynden-Bell}, 111

\bibitem[{Toomre} \& {Toomre}(1972)]{1972ApJ...178..623T}
{Toomre}, A., \& {Toomre}, J. 1972, \apj, 178, 623

\end{thebibliography}
%\begin{thebibliography}{99}
%% you can type \apj for ApJ, \aap for A&A, \apss for Ap&SS, etc. Please consult
%% the macro chjaa.cls. You can also find them in aasguide.tex (AASTeX for ApJ, AJ, PASP)
%% Please follow the format of ChJAA's reference list

%\end{thebibliography}

\label{lastpage}

\end{document}